\documentclass[preprint2]{aastex}

\newcommand{\kms}{\mbox{km\,s$^{-1}$}}
\newcommand{\pmu}{\mbox{mas\,yr$^{-1}$}}

\slugcomment{To appear in Astrophysical Journal Letters}

\shorttitle{Tidal Tails of Pal\,5}
\shortauthors{Odenkirchen et al.}

\begin{document}

\title{Detection of Massive Tidal Tails around the Globular Cluster \\
Pal\,5 with SDSS Commissioning Data}

\author{
Michael Odenkirchen\altaffilmark{1}, 
Eva K. Grebel\altaffilmark{1}, 
Constance M. Rockosi\altaffilmark{2},
Walter Dehnen\altaffilmark{1}, 
Rodrigo Ibata\altaffilmark{1},
Hans-Walter Rix\altaffilmark{1}, 
Andrea Stolte\altaffilmark{1},  
Christian Wolf\altaffilmark{1},
John E. Anderson Jr.\altaffilmark{3}, 
Neta A. Bahcall\altaffilmark{4},
Jon Brinkmann\altaffilmark{5},
Istvan Csabai\altaffilmark{6}, 
G. Hennessy\altaffilmark{7}, 
Robert B. Hindsley\altaffilmark{8}, 
\v{Z}eljko Ivezi\'{c}\altaffilmark{4}, 
Robert H. Lupton\altaffilmark{4},
Jeffrey A. Munn\altaffilmark{9}
Jeffrey R. Pier\altaffilmark{9}, 
Chris Stoughton\altaffilmark{3},
Donald G. York\altaffilmark{2}
}

\altaffiltext{1}{Max-Planck-Institut f\"ur Astronomie, K\"onigstuhl 17, 
D-69117 Heidelberg, Germany; odenkirchen@mpia-hd.mpg.de, 
grebel@mpia-hd.mpg.de}
\altaffiltext{2}{University of Chicago, Dept. of Astronomy \& Astrophysics, 
5640 South Ellis Ave., Chicago, IL~60637}
\altaffiltext{3}{Fermi National Accelerator Laboratory, P.O.~Box 500, 
Batavia, IL~60510}
\altaffiltext{4}{Princeton University Observatory, Princeton, NJ~08544; }
\altaffiltext{5}{Apache Point Observatory, P.O.\ Box 59, Sunspot, 
NM~88349-0059}
\altaffiltext{6}{Dept.\ of Physics \& Astronomy, The Johns Hopkins 
University, 3701 San Martin Drive, Baltimore, MD~21218}
\altaffiltext{7}{U.S.\ Naval Observatory, 3450 Massachusetts Ave.\ NW, 
Washington, DC~20392-5420  }
\altaffiltext{8}{Naval Research Lab, 4555 Overlook Ave. SW, Washington, 
DC~20375}
\altaffiltext{9}{U.S.\ Naval Observatory, Flagstaff Station, P.O. Box 1149, 
Flagstaff, AZ~86002-1149}

\begin{abstract}
We report the discovery of two well-defined tidal tails emerging from 
the sparse remote globular cluster Palomar 5. These tails 
stretch out symmetrically to both sides of the cluster in the direction 
of constant Galactic latitude and subtend an angle of $2.6^\circ$ on the 
sky.  The tails have been detected in commissioning data of the Sloan Digital 
Sky Survey (SDSS), providing deep five-color photometry in a 
$2.5^\circ$ wide band along the equator. The stars in the tails make up 
a substantial part ($\sim 1/3$) of the current total population of 
cluster stars in the magnitude interval $19.5 \le i^* \le 22.0$. This
reveals that the cluster is subject to heavy mass loss. The orientation 
of the tails provides an important key for the determination of the 
cluster's Galactic orbit. 

\end{abstract}


\section{Introduction}
Globular clusters are self-gravitating stellar
systems that experience a time-varying tidal potential as they orbit 
through their parent galaxy. 
Their dynamical evolution is driven by internal effects such as 
stellar evolution, two-body relaxation and binary heating, and by external 
effects induced by the galactic force field, i.e., heating by tidal shocks 
during disk and bulge passages and tidal stripping. 
Both internal and external effects should lead to a 
permanent loss of cluster members and to the eventual dissolution of the 
cluster. The Galactic globular clusters observed today are believed 
to be survivors from an initially much more numerous population. They 
appear to be in various stages of evolution and dissolution, depending 
on their initial conditions and their galactic orbits
(e.g., Chernoff \& Weinberg 1990, Djorgovsky \& Meylan 1994).
Numerical simulations predict that possibly as many as half of the 
present-day Galactic globulars will not survive for another 
Hubble time \citep{go97}. Observational confirmation of the gradual 
dissolution of globular clusters and determination of their mass loss 
rates is important in itself, but can also shed light on the 
formation history and structure of the Galactic halo and provide 
constraints on the Galactic potential.  

\begin{deluxetable}{ccl}
\tabletypesize{\footnotesize}
\tablecaption{Fundamental parameters of Pal\,5. \label{tab1}}
\tablewidth{0pt}
\tablehead{\colhead{Parameter} & \colhead{Value}   
& \colhead{Reference}}
\startdata
$\alpha,\delta$ (J2000)& $229.02^\circ, -0.11^\circ$ & \\
$l,b$ & $0.9^\circ, +45.9^\circ$ & \\
$d$ & 23.2~kpc & \citet{h99}\\
$M$ & $1.3\cdot 10^4 M_\odot$ & \citet{sh77} \\
$r_c$ & $2.9'$, 20~pc & \citet{tkd95} \\
$v_r$ & $-$56~\kms\ & \citet{sm85} \\
$\mu_{\alpha}\cos\delta,\mu_\delta$ & $(-2.44,-0.87)$~\pmu\ & 
\citet{scm93}\\
$\mu_{\alpha}\cos\delta,\mu_\delta$ & $(-1.0,-2.7)$~\pmu\ & 
\citet{rso98}\\
$\mu_{\alpha}\cos\delta,\mu_\delta$ & $(-2.55,-1.93)$~\pmu\ & 
Cudworth (priv.\ comm.)\\
\enddata
\end{deluxetable}

First signs of the existence of tidal debris around Galactic globular 
clusters were found in studies of globular cluster radial surface density 
profiles. In many cases the observed profiles deviate from the profile
of a best-fit King model at the outermost radii and extend beyond the 
conventional limiting radius set by the model (e.g., Grillmair et al.\ 1995; 
Lehmann \& Scholz 1997).  
Since the debris produced by a dissolving globular cluster will have 
similar orbital parameters as the cluster itself it is expected to drift 
away from the cluster in a leading and a trailing tail which are aligned 
with the cluster's orbit. 
Searches for such tidal tails in the vicinity of Galactic globular 
clusters require large area coverage and have so far been conducted with 
photographic material \citep{grill95,lmc00,testa00}. 
While spatially distinct star count overdensities were detected around 
many clusters, density fluctuations caused by distant galaxy clusters, 
variable foreground reddening or photographic plate inhomogeneities have 
been found to be serious contaminants and have left the location and shape 
of the tidal tails in most cases uncertain.

The Sloan Digital Sky Survey (SDSS; see \citet{york00, gunn98, fu96}) 
is designed to provide homogeneous, deep 
($R \sim 23$) CCD imaging in five passbands ($u',g',r',i',z'$) with 
contiguous coverage of $10^4\Box^\circ$  in the Northern Galactic Cap.  
Such a high-quality database is ideally suited for studies of Galactic 
structure (e.g., Yanny et al.\ 2000; Ivezic et al.\ 2000)
and lends itself in particular to searches for tidal debris 
with unprecedented depth and spatial coverage. 

The SDSS commissioning scans happen to cover Palomar 5, a remote, 
sparse halo cluster. 
The peculiar characteristics of this cluster, i.e., very low 
mass, large core radius, and low central concentration (see Tab.\ 1) 
suggest that Pal\,5 may have lost a significant amount 
of its initial mass and perhaps be close to disruption.
Its radial velocity combined with determinations of its absolute 
proper motion suggest an eccentric orbit with passages through the 
Galactic disk inside the solar circle \citep{rso98}. 

In view of these indications we have searched the SDSS data for tidal 
debris around Pal\,5. In Section 2 we describe our search method. 
Results are presented in Section 3.

\begin{figure*}[t]
\begin{center}

\includegraphics[scale=0.9,bb=42 505 565 675,clip=true]{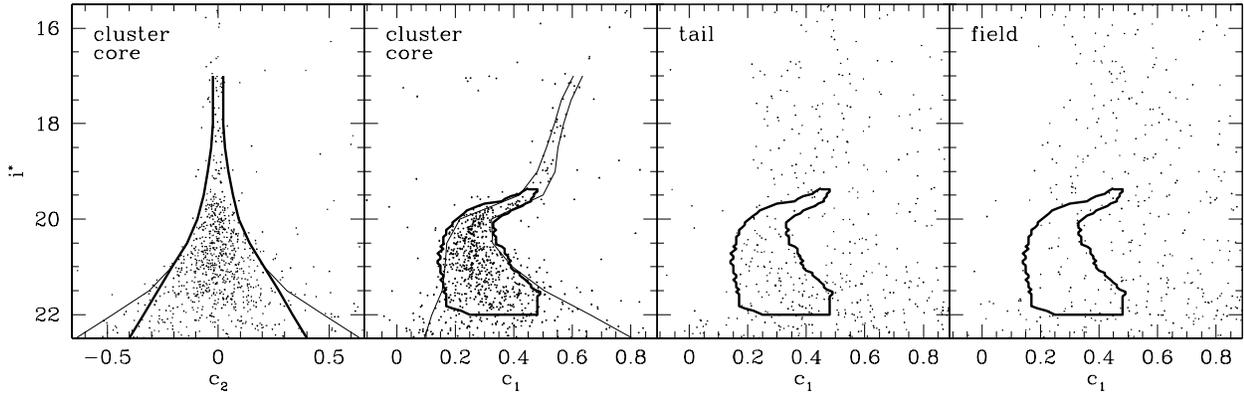}
\figcaption[pal5fig1.eps]{Color-magnitude diagrams for stellar SDSS 
sources in the field of Pal\,5.
From left to right: 1. $(c_2,i^*)$ for objects within $3'$ of the 
cluster center (core).
2. $(c_1,i^*)$ for objects within $3'$ of the cluster center (core).  
3. $(c_1,i^*)$ for objects in two circles centered on the cluster tails 
4. $(c_1,i^*)$ for objects in two circles in the outer field
Note that the field sizes of the tail and field samples shown here 
are $8\times$ that of the core sample. 
The thick lines indicate our photometric selection criteria for 
cluster member candidates, the thin lines give $2\sigma$-limits for 
the expected dispersion of cluster stars in $c_1$ and $c_2$ due to 
photometric errors.  
\label{fig1}}

\end{center}
\end{figure*}

\section{Search method}

\begin{figure*}[t]
\begin{center}

\includegraphics[scale=0.63,angle=270,bb=40 45 275 770,clip=true]{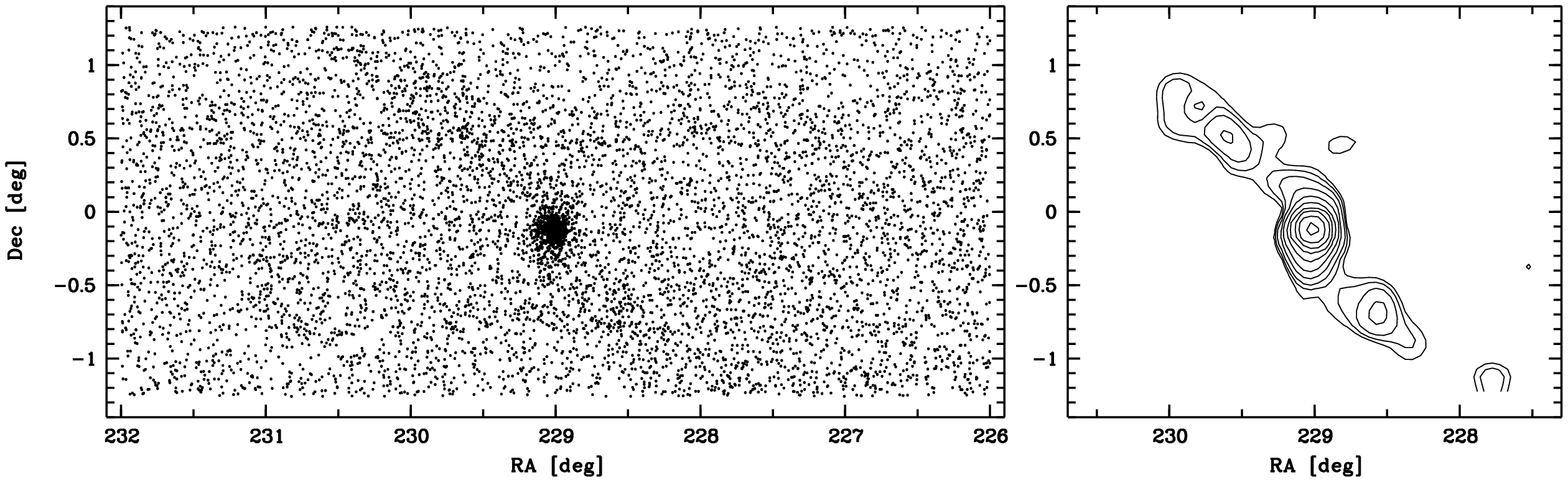}
\figcaption[pal5fig2.eps]{Projected spatial distribution of 
the photometric cluster member candidates selected with the method described 
in Section 2. Left: individual star positions. Right:
contours of the surface density derived by adaptive 
kernel estimation. Contour levels are 0.19, 0.21, 0.24, 0.3, 0.4, 0.6, 1.0, 
2.0, 4.0 and 8.0 stars/$\Box '$. The background has a mean of 0.13 
stars/$\Box '$. \label{fig2}}

\end{center}
\end{figure*}

In the current phase of SDSS, the region of Pal\,5 is covered by 
an equatorial band with declinations $-1.25 \le \delta \le +1.25$. Our
investigation concentrates on the range $226^\circ \le \alpha \le 232^\circ$,
i.e., $3^\circ$ east and west of the cluster, and on objects classified 
as SDSS point sources. 
The typical accuracy of the relative photometry in this data set is 
better than 0.03~mag for stars brighter than 20.0~mag in $g^*$, $r^*$, 
or $i^*$, and is $\sim 0.1$~mag at $u^*= 21.0, g^*= 22.2, r^*= 21.8, 
i^*= 21.4$, and $z^* = 20.0$ mag ($^*$ is used to indicate that the 
photometric zeropoints are still preliminary). 
In order to enable corrections for interstellar extinction the SDSS data 
provide individual extinction values based on the maps of Schlegel et 
al.\ (1998). These values (which are small in the field of Pal5) have 
been subtracted from the observed stellar magnitudes.  

One straightforward method for tracing the spatial distribution 
of cluster stars in a large field around a cluster is as follows: 
(i) Define the distribution of the cluster population in color-magnitude 
space by means of the stars near the cluster center, and that 
of the field stars by stars from a sufficiently distant part of the field. 
(ii) Determine the significance of each point in color-magnitude space 
by means of the local signal-to-noise ratio $s$ of the expected true number 
of cluster stars, 
(i.e.\ the counted number of cluster stars minus the estimated number 
of contaminating field stars, see eq.(2)) in the surroundings of this point.  
(iii) Select those regions in color-magnitude space with high significance, 
i.e., with local signal-to-noise $s$ above a certain threshold, 
extract from the entire sample all stars which lie in these particular
color-magnitude regions, and analyze the spatial configuration of these
stars.

This way of empirical photometric filtering was first described and 
used in the two-color study by \citet{grill95}. It could, 
in principle, be easily generalized to a higher dimensional photometric 
space and thus be directly applied to the multimagnitude or 
multicolor-magnitude data of SDSS. 
However, in the case of the very sparse cluster Pal\,5 it 
turned out that the number of cluster stars was too small to provide 
reasonable star count statistics in multidimensional color-magnitude 
cells. In order to avoid problems with noise from small numbers      
we thus worked with projections on color-magnitude planes, viz. 
those spanned by the magnitude $i^*$ and the color indices $u^*-g^*, 
g^*-r^*, r^*-i^*$, and $i^*-z^*$ or orthogonal combinations of the latter.

From $g^*-r^*$ and $r^*-i^*$ we tailored new color indices $c_1,c_2$ as
given in eq.\ (1ab), 
with the advantage that for cluster stars the systematic variation of color 
with magnitude is mostly contained in $c_1$ while variations in $c_2$ 
are mostly random, i.e.\ due to natural spread and observational errors 
(see left panels of Fig.1). 

\begin{mathletters}
\begin{eqnarray}
c_1&=& 0.907 (g^*-r^*) + 0.421 (r^*-i^*)\ \\
c_2&=& -0.421 (g^*-r^*) + 0.907 (r^*-i^*)\  
\end{eqnarray}
\end{mathletters}

The indices $u^*-g^*$ and $i^*-z^*$ were considered separately 
because they are less accurate than the others due to decreased sensitivity 
in the bluest and reddest photometric band.   

As the first step of photometric filtering we applied simple 
magnitude-dependent color cuts in $u^*-g^*$, $i^*-z^*$, and $c_2$ 
according to the borders of the cluster population defined by stars 
from the cluster core ($r\le 3'$), and according to the estimated spread 
due to photometric errors (for $c_2$ cf.\ leftmost panel of Fig.\ 1). 

After this preselection, we used Grillmair's method for filtering 
in the color-magnitude plane of $(c_1,i^*)$. The local signal-to-noise 
$s$ of the expected true number of clusters stars was determined on a 
grid of mesh size 0.01~mag in $c_1$ and 0.05~mag in $i^*$. For each 
grid point (index $k$) the number $n_c(k)$ of stars in a circle of 
radius $12'$ around the cluster center and the number $n_f(k)$ of stars 
at more than $2^\circ$ angular distance from the cluster center were 
counted in a color-magnitude box of width 0.09~mag in $c_1$ and 0.35~mag 
in $i^*$ centered on that point. With $n_c$ representing the number of 
cluster stars plus underlying field stars, $n_f$ the number of field stars, 
and $q$ the ratio of the areas on which $n_f$ and $n_c$ have been sampled, 
$s$ was calculated as given by eq.(2). 
      
\begin{eqnarray}
s(k)=\frac{n_{c}(k)-q^{-1}n_f(k)}{\sqrt{n_{c}(k) + q^{-2}n_f(k)}}
\end{eqnarray}

The size of the color-magnitude boxes and the overlap between boxes 
around neighboring grid points assure that $s$ is a sufficiently smooth 
function. From the local signal-to-noise $s$ one obtains a
filtering mask in the $(c_1,i^*)$ plane by setting a threshold 
$s_{lim} < s_{max}$ and by isolating the region in the 
grid (around the maximum of $s$) with $s \ge s_{lim}$. In order to find 
the most appropriate mask, we went through a series of gradually 
decreasing thresholds, counted for each threshold the cumulative number 
of stars in the corresponding mask in the area of the cluster's tails 
($N_t$) and in the outer field ($N_f$), and determined 
from these numbers the cumulative signal-to-noise ratio SNR of the expected 
true number of cluster stars in the area of the tails (eq.(3)).  
 
\begin{eqnarray}
{\rm SNR} = \frac{N_t - w^{-1}N_f}{\sqrt{N_t + w^{-2}N_f}}
\end{eqnarray}

The filtering mask was then chosen such that the cumulative signal-to noise 
reaches a maximum. As shown in Fig.~1 (second panel from left) 
this mask cuts out the zone from the bottom of the subgiant branch 
to the main-sequence turnoff and further down the main sequence 
to $i^*= 22.0$ mag. In the range $19.5 \le i^* \le 21.5$ the 
width of the mask approximately coincides with the $2\sigma$ limits for 
the dispersion of cluster stars in $c_1$ as derived from the median 
values of the estimated photometric errors. The two panels
on the right in Fig.~1 give an example of the detection of cluster 
member candidates outside the cluster using the filter mask in the area 
of the cluster's tails and in the area of the outer field. 

The spatial configuration of the complete sample of member 
candidates obtained by the photometric filtering process  
is shown in Fig.\ 2 (left panel). 
In the final step of data processing,
the distribution of individual star positions was transformed 
into a smooth surface density function by means of adaptive 
kernel estimation \citep{sv86}. 
A standard parabolic kernel was used, with the kernel radius set  
to the angular distance of the 70th nearest neighbor of each star.
This yields the surface density distribution shown in the 
contour plot of Fig.\ 2 (right panel).

\section{Discussion}
\subsection{Characteristics of the tails} \label{characteristics}
Fig.\ 2 shows that the density enhancements of point sources 
around the cluster form two spectacular tails which emerge from 
the cluster in northern and southern direction and turn over
to the north-east and south-west, respectively,  at angular distances of 
$\sim 0.2^\circ$ (80~pc in projected linear distance) from the cluster 
center. 
The tails stretch out almost symmetrically to both sides and exhibit 
clumps at a distance of $\sim 0.8^\circ$, i.e., 320~pc from the 
cluster. In total, the tails are visible along an arc of $2.6^\circ$. 
A weaker clump at the southern edge of the current field suggest that the 
tails might in fact continue to even larger distances.  

In the two big clumps, the surface density of stars that fall inside our 
colour magnitude filter is about 2.3 times as high as in the surrounding 
field. 
Summing up the number of stars above background in the region of the tails 
and comparing them to the stars within a circle of radius $r < 12'$ around 
the cluster center we find that within our color-magnitude window the tails 
contain $\sim 0.48$ times the number of stars in the cluster. 
In other words, the tails comprise $\sim 32\%$ of the currently detected 
total population of cluster stars at and below the main-sequence turnoff. 
This is a rough 
(but conservative) estimate because the object is seen in a 
non-face-on projection which does not reveal a clear border between cluster 
and tail. 
Nonetheless it gives impressive evidence for heavy mass loss, confirming 
conjectures drawn from the low mass and low concentration of the cluster.    

The structure of the observed tails follows the principal 
expectations for tidal tails and closely agrees with the results 
of recent N-body simulations of globular cluster tides 
\citep{clm99}. Basically,
cluster members drift to the outer part of the cluster after acceleration 
in disk or bulge shocks and leave the cluster in the vicinity of 
the (Lagrange-) points of force balance between the cluster and the tidal 
field, i.e., in the direction to the galactic center and anticenter. 
Due to differential galactic rotation their trajectories then 
bend around and continue approximately parallel to the orbit of 
the cluster. The appearance of clumps in the tails is also 
supported by numerical simulations.
They can either be caused by the enhanced release of particles after 
strong shock events or by caustics of the trajectories in phase space.

\subsection{The role of contaminants} \label{contamination}
In view of the striking resemblance of the detected structures to
the expected properties of tidal tails, significant contamination by 
clustered background objects seems a priori unlikely. Nevertheless 
we checked this point by analysing the density and colors of non-pointlike
SDSS sources around Pal\,5 which by their shape are classified as 
galaxies. 
Their spatial distribution is clumped and reveals known galaxy 
clusters like Abell\,2050 and 2035. 
However, most of these sources do not fall into the color-magnitude 
window for members of Pal\,5.
If our selection criteria for Pal\,5 members are applied to 
the galaxy sample, the surface density of galaxies drops to the level of 0.3 
to 0.6 of the field background density of the stellar sample. 
Moreover, the pattern of density variations in the galaxy sample does 
not correlate with the location of the tidal tails.    
Therefore, objects like those in the galaxy sample are not likely
to cause significant disturbances in the stellar sample.  
The only remaining contaminants are compact galaxies with bluer 
colors that may not be well represented in the sample of known galaxies.
We believe that such objects have mostly been eliminated by 
our color cut $u^*-g^* \le 0.4~mag$. Finally,
fluctuations in stellar surface density due to variable interstellar 
absorption can be ruled out because the mean level of absorption 
in the region around Pal\,5 is low and there is no hint for 
strong variations (values of $E_{B-V}$ in the maps of Schlegel et 
al.\ (1998) range between 0.05~mag and 0.07~mag).

\begin{figure}[t]
\begin{center}

\includegraphics[scale=0.57,angle=270,bb=45 60 350 425,clip=true]{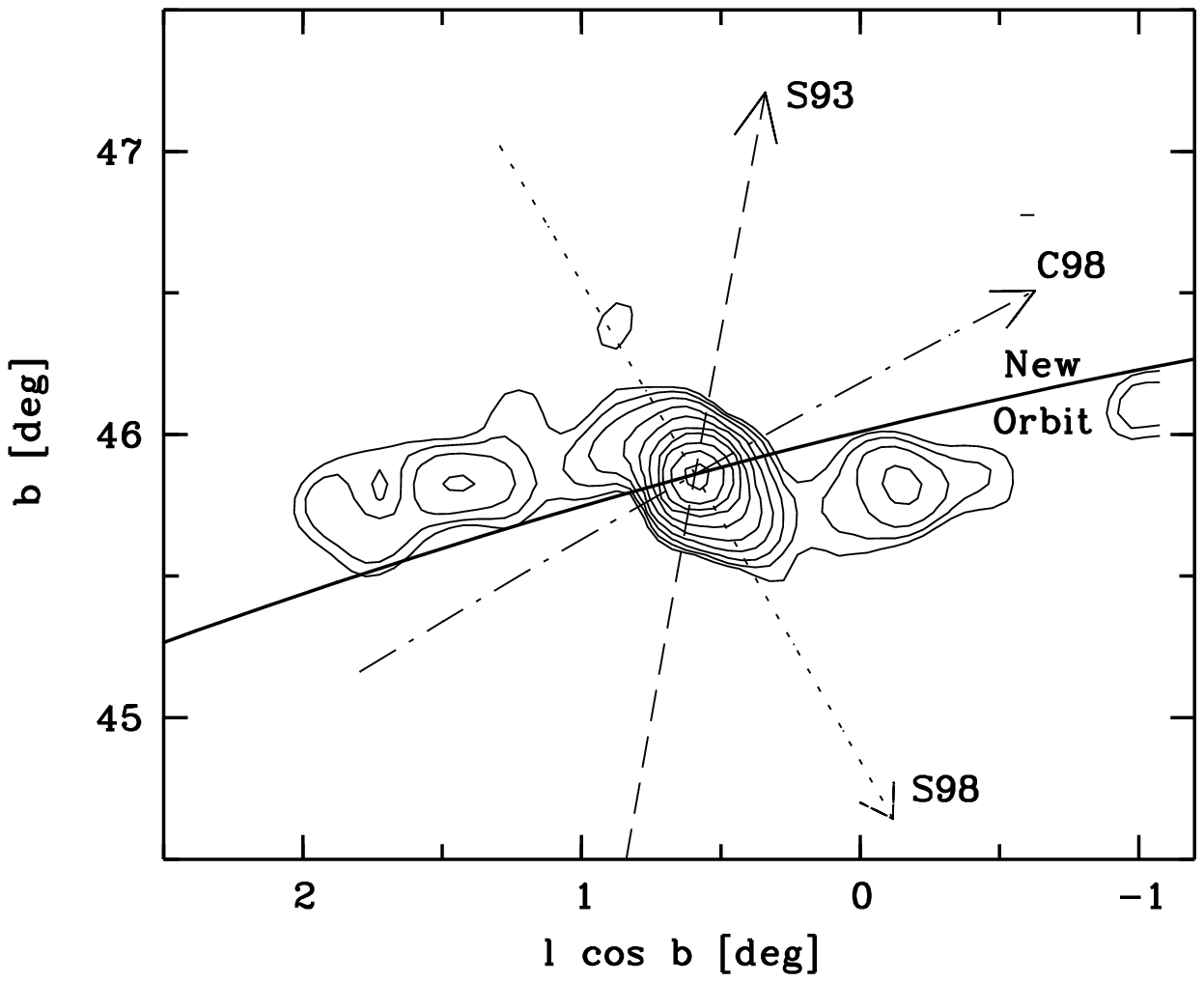}

\figcaption[pal5fig3.eps]{Contour plot of the surface density of 
cluster candidates in galactic coordinates ($l\cos b, b$) overlayed 
with different orbital paths of the cluster according to different 
determinations of its absolute proper motion:
S93 = \citet{scm93}; S98 = \citet{rso98}; 
C98 = Cudworth (unpublished), cited in \citet{rso98}.  
The solid line presents our improved estimate of the orbit 
based on the geometry and orientation of the tidal tails 
(fixing the direction of tangential motion) and the proper motions
by Cudworth and by Scholz et al.\ (estimating the tangential 
velocity).  \label{fig3}}
\end{center}
\end{figure}

\subsection{Implications for and from the cluster's orbit} \label{orbit}

The orientation of the tails provides unique information on the 
direction of the cluster's tangential motion since it is known that 
the leading and trailing parts must fit in with the inner 
and outer side of the local orbit, respectively. 
Fig.\ 3 reveals that the tails stretch out in the direction of constant $b$, 
and that the tangential motion is very likely westward 
(prograde rotation). The absolute proper motions by \citet{scm93}
and \citet{rso98} yield very different predictions for the local 
orbit, although at least the sense of rotation is in agreement. 
The proper motion given by Cudworth (priv.\ comm., unpublished revision 
of the work by Schweitzer et al.) yields a local orbit which lies 
closer to the tails. From the observed orientation of the tails we 
estimate that the tangential velocity vector points $\sim 15^\circ$
north of the line of constant $b$. In order to meet this constraint 
the proper motion needs to be modified by not more than 0.4~\pmu\ 
w.r.t.\ Cudworth's values. 
We thus adopt $\mu_l\cos b,\mu_b = -0.93,+0.25$\,\pmu\ for the cluster's 
proper motion in the galactic rest frame. With these values   
we obtain an orbit (using the 
galactic potential of Allen \& Santillan 1991) with apo- and perigalactic 
distances of 19.0 kpc and 7.0~kpc, and with disk passages at 
$-$137~Myr, $-$292~Myr and $-$472~Myr, taking place at Galactocentric 
radii of 9.4, 18.4 and, 8.3~kpc, respectively. 
Similar orbits are obtained with the more detailed galactic potentials
of \citet{db98}. 
We tend to believe that the observed overdensities close 
to the cluster result from  
the latest disk passage while the clumps in the tails at distances of 
$\sim 0.8^\circ$ from the cluster might be associated with the 
earlier passage through the inner disk about 470~Myr ago. This, however, 
has to be investigated more thoroughly with N-body simulations. 
The next passage through the galactic disk predicted by our model orbit 
will be in $113$~Myr and will happen close to perigalacticon. 
If true, this will again produce a strong tidal shock that may 
eventually dissolve the cluster completely.     

\subsection{Outlook} \label{outlook}

The SDSS will eventually cover a much larger region around Pal\,5 than 
currently available. Larger area coverage will enable us to constrain the 
orbit and mass loss more tightly. We can further constrain our model 
orbit by obtaining radial velocities of stars in the tails. We predict 
a local radial velocity gradient of $5.7$~\kms\,deg$^{-1}$, 
i.e.\ $\sim 9$~\kms\ difference between the radial velocities of stars 
in the two tidal tail clumps. Since Pal\,5 contains very few luminous 
red giants, even fewer are expected in its tails, thus kinematic studies 
will have to concentrate on fainter stars requiring large telescopes.

\acknowledgments
The Sloan Digital Sky Survey (SDSS) is a joint project of The University of 
Chicago, Fermilab, the Institute for Advanced Study, the Japan
Participation Group, The Johns Hopkins University, the Max Planck Institute 
for Astronomy, New Mexico State University,
Princeton University, the United States Naval Observatory, 
and the University of Washington. 
Apache Point Observatory, site of the SDSS, is operated by the Astrophysical
Research Consortium. 
Funding for the project has been provided by the Alfred P.\ Sloan Foundation, 
the SDSS member institutions, the National Aeronautics and Space 
Administration, the National Science Foundation, the U.S. Department of 
Energy, and Monbusho. The SDSS Web site is \url{http://www.sdss.org/}.




\clearpage





\end{document}